\documentstyle[11pt,epsfig,color]{article}
\begin{document}

\begin{center}
{\Large\bf Kinematical models of latetime cosmology and the statefinder diagnostic}
\\[15mm]
Ankan Mukherjee\footnote{Email: ankan.ju@gmail.com}
\vskip 0.5 cm
{\em $^{1}$Centre for Theoretical Physics,\\ Jamia Millia Islamia, Jamia Nagar, New Delhi-110025, India}\\[15mm]
\end{center}

\vspace{0.5cm}
\vspace{0.5cm}
\pagestyle{myheadings}
\newcommand{\be}{\begin{equation}}
\newcommand{\ee}{\end{equation}}
\newcommand{\bea}{\begin{eqnarray}}
\newcommand{\eea}{\end{eqnarray}}

\begin{abstract}
The present work deals with kinematical models of latetime cosmology. It is based on purely phenomenological assumption about the deceleration parameter. The models are confronted to observational data sets of type Ia supernovae distance modulus measurements and measurements of Hubble parameter at different redshift. Constraints on the cosmological parameters are obtained by Markov Chanin Monte Charlo (MCMC) analysis using the observation data sets. The values of present Hubble parameter, deceleration parameter and the redshift of transition from decelerated to accelerated phase of expansion are estimated for the present kinematical models. Further the properties of dark energy for the present models are explored with general relativistic assumptions. The dark energy diagnostics, namely the {\it Om} diagnostic and the {\it statefinder} are adopted for a comparison of the present models. A phase space constructed of two different statefinder parameters breaks the degeneracy of the models. It is observed that the kinematical models attain the corresponding $\Lambda$CDM value on the phase space in the course of evolution.  The evolution of matter density contrast at linear level has also been studied for the present kinematical models. 
\end{abstract}

\vskip 1.0cm


\section{Introduction}

The observed phenomenon of cosmic acceleration \cite{Riess:1998cb,Perlmutter:1998np,Schmidt:1998ys} is still an enigma for cosmologists. There are two different direction of finding a theoretical explanation of the alleged accelerated expansion of the universe. Within the framework of General Relativity (GR), the cosmic acceleration can be explained by introducing an exotic component in the energy budget of the universe. It is dubbed as {\it dark energy}. Dark energy with its characteristic negative pressure can generate the accelerated expansion. The other way to look for a possible explanation cosmic acceleration is the modification of the theory of General Relativity. It is not yet been ascertained  whether dark energy or the space-time geometry itself is responsible for the accelerated expansion. But GR is highly successful to explain the local astronomical and cosmological observations than the modified gravity theories.

Though the dark energy cosmology is efficient to explain cosmological observations, there is hardly any certain knowledge about physical entity of dark energy. Various theoretical prescriptions regarding dark energy are there in the literature. The cosmological constant or vacuum energy density \cite{Padmanabhan:2002ji,Peebles:2002gy}, scalar field models of dark energy like quintessence \cite{Tsujikawa:2013fta}, k-essence \cite{Scherrer:2004au}, tachyon field \cite{Bagla:2002yn} etc, fluid model of dark energy like chaplygin gas model \cite{Bento:2003dj}, are amongs them. Different theoretical aspects of dark energy are comprehensively reviewed by Copeland, Sami and Tsujikawa \cite{Copeland:2006wr}.  The cosmological constant ($\Lambda$) model of dark energy along with cold dark matter (CDM) is well consistent with most of the cosmological observations. Hence it is accepted as the standard cosmological paradigm, also dubbed as {\it concordance cosmology}. However, there are certain issues that urge to look for alternatives of cosmological constant model. One important theoretical issue is the humongous discrepancy between the observationally estimated value of cosmological constant and the value of vacuum energy density calculated in quantum field  theory \cite{Padmanabhan:2002ji}. The other astonishing fact is the same order of magnitude value of cosmological constant and the matter energy  density at present epoch.  It is called the {\it cosmic coincidence problem}. Due to these theoretical issues, time-evolving dark energy have gained attention in this context. Some recent cosmological observations like the Lyman-$\alpha$ forest BAO measurement of Hubble parameter at redshift $2.34$ by Baryon Oscillation Spectroscopic Survey (BOSS) \cite{Font-Ribera:2013wce} and the local measurement of Hubble constant ($H_0$) by Hubble Space Telescope (HST) \cite{Riess:2016jrr,Riess:2019cxk} are in disagreement with concordance cosmological model the $\Lambda$CDM. Other dark energy models are also not very successful to alleviate these disagreements. The local measurements are found to be not very sensitive to background cosmological models \cite{Dhawan:2020xmp}.

Reconstruction of cosmological model is a reverse engineering based on cosmological observations. The idea is to figure out the evolution of certain cosmological quantities from observational data in parametric or non-parametric fashion. Aspects of reconstruction of dark energy model have been comprehensively reviewed by Sahni and Starobinsky \cite{Sahni:2006pa}. The model can be reconstructed with some prior assumption about the dynamics of dark energy. Another way of reconstruction of cosmological model is in kinematic approach. Kinematic approach to reconstruct cosmological models are investigated in the present work. A kinematic approach of reconstruction only assumes the homogeneity and isotropy of the universe at cosmological scale. It is independent of any assumption about the dark energy model and even the theory of gravity. A kinematic approach to reconstruct cosmological evolution  using a Taylor expansion of the Hubble parameter has been discussed by Mukherjee {\it et al} \cite{Mukherjee:2018oll}. A model independent approach to constraint the kinematics of late-time cosmology has been discussed by  Shafiello {\it et al} \cite{ShafielloKimLinder} and by Haridasu {\it et al} \cite{Haridasu:2018gqm}. Parameterizations of deceleration parameter in the context of late-time cosmology has been discussed by Gong and Wang \cite{Gong:2006tx,Gong:2006gs,Mamon:2018dxf}. Campo {\it et al} \cite{delCampo:2012ya} discussed parameterizations of deceleration parameter based on thermodynamical consequences. Kinematical reconstruction using higher order kinematic terms are discussed by Rapetti {\it et al} \cite{Rapetti:2006fv}, by Zhai {\it et al} \cite{Zhai:2013fxa}, and by Mukherjee and Banerjee \cite{Mukherjee:2016trt,Mukherjee:2016shl}. Dark energy cosmology with equivalent descriptions and cosmographical test has been reviewed by Bamba {\it et al} \cite{Bamba:2012cp}. A few more kinematical and cosmographical analysis of late-time cosmic evolution are referred there in \cite{Aviles:2012ay,Gruber:2013wua,Capozziello:2017nbu,Wang:2009fy}.

In the present work, the late-time cosmological evolution is studied through phenomenological parameterizations of the deceleration parameter. Deceleration parameter is the dimensionless kinematical parameter that contains the second order time derivative of the scale factor. Deceleration parameter is a measure of cosmic acceleration in a dimensionless way. It is important to check whether redshift of transition from decelerated to accelerated phase varies significantly with variation in the model or it remains stable. The transition redshift is considered as a free parameter in the present analysis to probe this issue. The kinematical models are independent of any prior assumption about the dark energy. The nature of dark energy equation of state for the kinematical models are studied based on GR assumption about the distribution of the components in the energy budget. Dark energy diagnostics, namely the {\it Om} and the {\it statefinder} are excellent in breaking the degeneracy of dark energy models. {\it Om} and {\it starefinder} parameters are invoked to break the degeneracy in the kinematical models and also to compare  the models with $\Lambda$CDM scenario. The other important aspect of the present study is the matter density perturbation for the kinematical models which is again based on general relativistic equation of matter density perturbation. The evolution of matter density perturbation at linear level is studied for the reconstructed kinematical models.

\par The paper is organized as the following. In section (section \ref{para}), reconstruction of latetime cosmological model from the phenomenological parameterizations of deceleration parameter is discussed. The statistical analysis of the models using cosmological observations and the constrains on kinematical parameters are presented in section \ref{Stat}. Evolution of different cosmological parameters and dark energy equation of state are also discussed in section \ref{Stat}. In section \ref{statefinder}, the different diagnostics of dark energy models are investigated. The evolution of matter density perturbation is studied in section \ref{deltam}. Finally in section \ref{conclu}, it is concluded with a overall discussion about the result.

\section{Reconstruction of the kinematical models}
\label{para}

\begin{figure}[tb]
\begin{center}
\includegraphics[angle=0, width=0.68\textwidth]{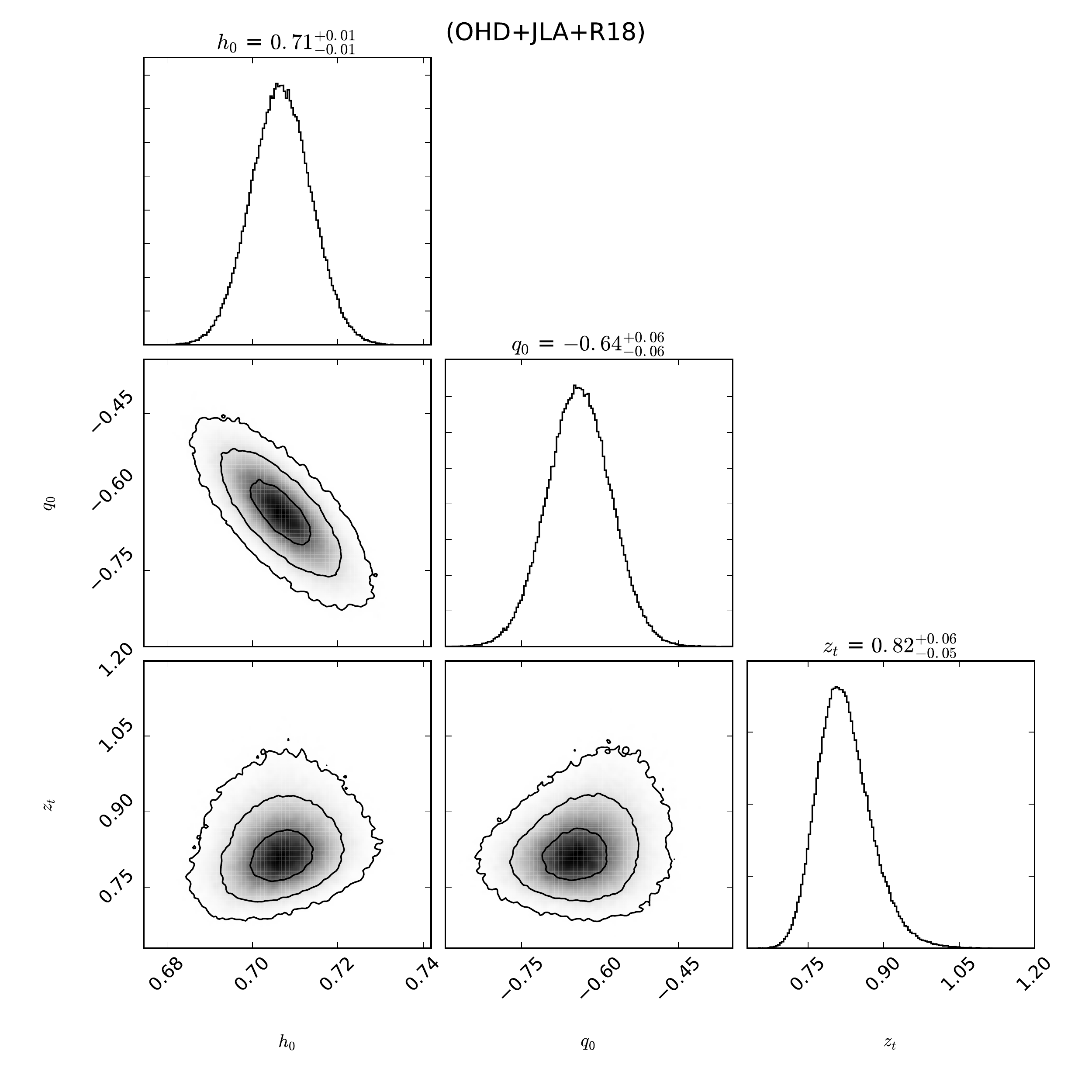}
\end{center}
\caption{{\small Marginalize posterior distributions and the 2D confidence contours of the parameters ($h_0,q_0,z_t$) for the reconstructed kinematical Model I, obtained in the analysis combining OHD, JLA and R18. The 1$\sigma$, 2$\sigma$ and 3$\sigma$ contours on 2D parameter spaces are shown. The best fit values of the parameters and the associated 1$\sigma$ uncertainties are also mentioned.}}
\label{cont_M1}
\end{figure}

\begin{figure}[tb]
\begin{center}
\includegraphics[angle=0, width=0.68\textwidth]{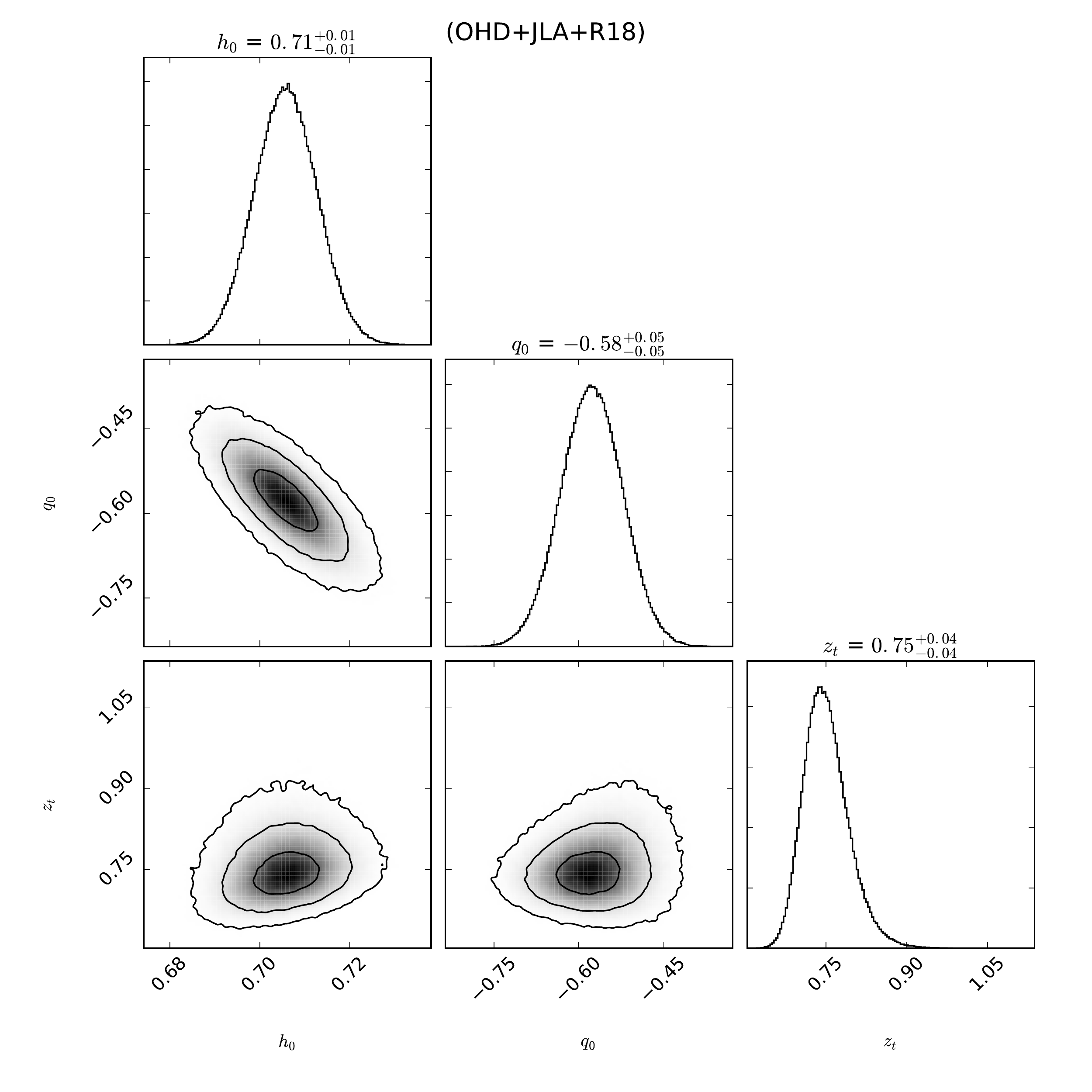}
\end{center}
\caption{{\small Marginalize posterior distributions and the 2D confidence contours of the parameters ($h_0,q_0,z_t$) for the reconstructed kinematical Model II, obtained in the analysis combining OHD, JLA and R18. The 1$\sigma$, 2$\sigma$ and 3$\sigma$ contours on 2D parameter spaces are shown. The best fit values of the parameters and the associated 1$\sigma$ uncertainties are also shown.}}
\label{cont_M2}
\end{figure}

\begin{figure}[tb]
\begin{center}
\includegraphics[angle=0, width=0.68\textwidth]{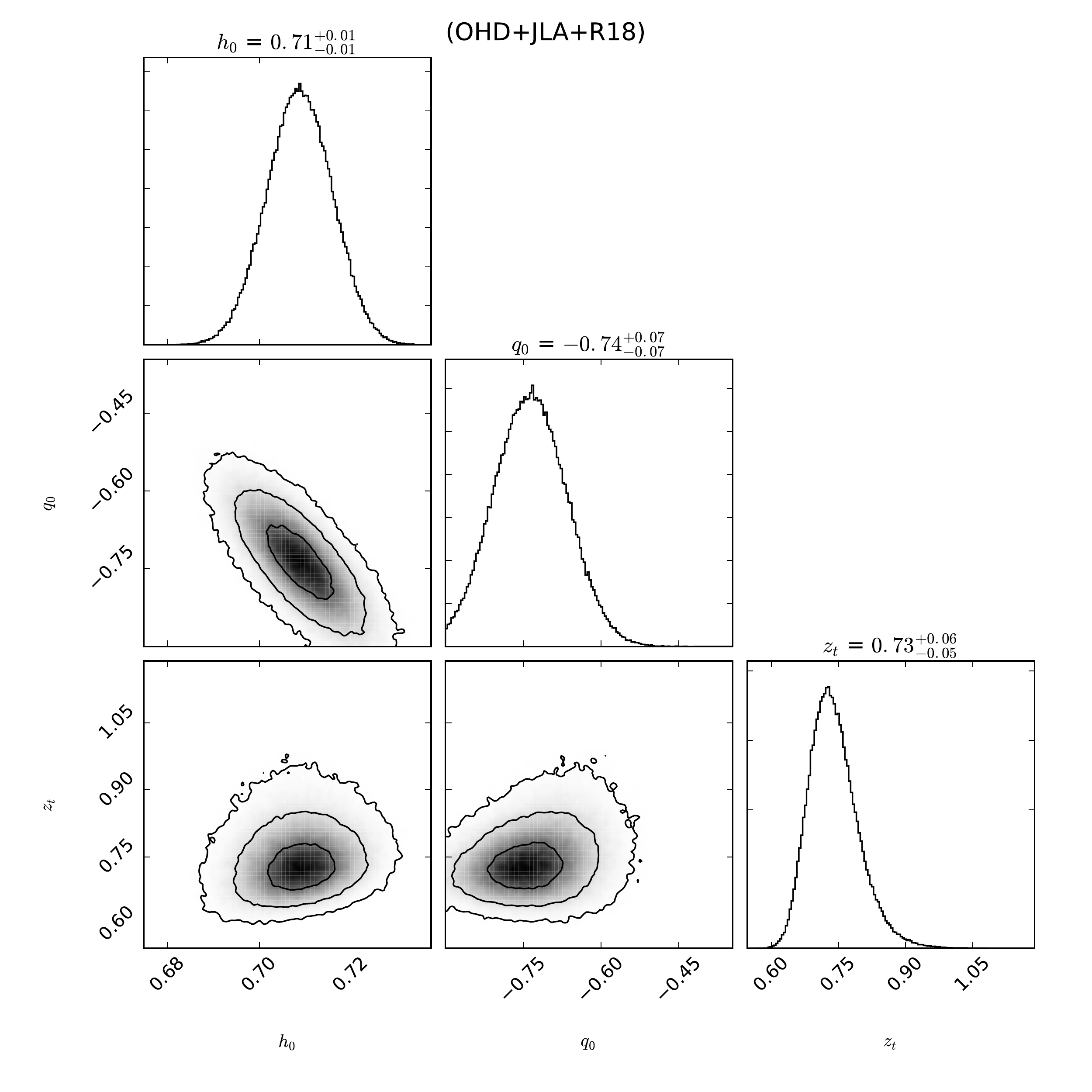}
\end{center}
\caption{{\small Marginalize posterior distributions and the 2D confidence contours of the parameters ($h_0,q_0,z_t$) for the reconstructed kinematical Model III, obtained in the analysis combining OHD, JLA and R18. The 1$\sigma$, 2$\sigma$ and 3$\sigma$ contours on 2D parameter spaces are shown. The best fit values of the parameters and the associated 1$\sigma$ uncertainties are also shown.}}
\label{cont_M3}
\end{figure}

The kinematic approach to the reconstruction of cosmological model is purely based on the assumptions of the {\it cosmological principle}, that is the universe is spatially homogeneous and isotropic. In a kinematic approach, the parameters defined in terms of the {\it scale factor} ($a(t)$) and its time-derivatives are utilized to reconstruct the model. It is independent of any prior assumption about the physical nature of dark energy, the distribution of different components in the energy sector and even any assumption about the gravity theory. 

\par The first order kinematic term is the Hubble parameter, defined as, $H(t)=\frac{\dot{a}}{a}$, where the overhead dot denotes differentiation with respect to time $t$. Hubble parameter gives the expansion rate at cosmological scale. It is convenient to use the redshift as the argument instead of cosmic time. Redshift is defined as $z=-1+\frac{a_0}{a(t)}$, where $a_0$ is the present scale factor. The Hubble parameter can also be presented as a function of redshift.  The second order kinematic term, which is the measure of the cosmic acceleration in a dimensionless way, is the deceleration parameter.  It is defined as $q(t)=-\frac{\ddot{a}/a}{H^2}$. This can be written in terms of Hubble parameter and its derivative with respect to the redshift $z$ as,
\be
q(z)=-1+(1+z)\frac{H'}{H},
\label{qz}
\ee
where the ``prime" denotes the differentiation with respect to $z$. A positive value of the deceleration parameter indicates the decelerated expansion of  the universe and a negative deceleration parameter indicates an accelerated expansion. It has been confirmed by cosmological observation that the universe was going through a decelerated phase of expansion in the past and presently it is in a phase of accelerated expansion. It has also been assured that the transition from decelerated to accelerated phase of expansion happened in recent past \cite{Riess:2004nr}. In the present analysis, we have adopted three different parameterizations of deceleration parameter. These parameterizations are purely phenomenological and motivated from the observational facts. These parameterizations are given as, Model I. $q(z)=q_0+q_1\frac{z}{(1+z)}$, Model II. $q(z)=q_0+q_1\frac{z(1+z)}{1+z^2}$, Model III. $q(z)=q_0+q_1\left[1-\frac{1}{(1+z)^2}\right]$. The utility of reconstructing kinematical parameters, like the deceleration parameter, is that kinematics of cosmic evolution, obtained in the analysis, does not depend on the specific gravity theory. On the other hand, the disadvantage is that we can not constrain the properties of the source of cosmic acceleration. Though the parametrizations of $q(z)$ are purely phenomenological, the phases of cosmic acceleration are taken care of in the parametrizations. For all these three expressions at $z=0$, $q(z=0)=q_0$. The observational constraints on $q_0$ would provide a measurement of present cosmic acceleration without any assumption about the dark energy properties. Cosmological observations of large scale structure have confirmed that the universe went through a phase of decelerated expansion in the past which was mostly matter dominated. In a matter dominated universe, the deceleration parameter remains constant with a value of $0.5$. In the expressions of $q(z)$, utilized in the present analysis, we see that the $q(z)$ attain a constant value $q\sim (q_0+q_1)$ at higher redshift.

Another important kinematical quantity is the redshift of transition from decelerated to accelerated phase of expansion ($z_t$) where the $q(z_t)=0$. In the present analysis the constraint on $z_t$ are also obtained in for the present kinematical models. One of the prime endeavour of introducing three different parametrizations of $q(z)$ is to check whether the value of present deceleration parameter ($q_0$) and the transition redshift ($z_t$) remain consistent or vary significantly with the change in the expression of deceleration parameter. The variation of higher order kinematical terms and the state statefinder parameters for different expressions of $q(z)$ are also investigated in the present study.

Equation (\ref{qz}) shows that the first integral of $q(z)$ will give the expression of Hubble parameter. For these models, the expressions of Hubble parameter are obtained as,

\be
Model ~ I.~~~H(z)=H_0(1+z)^{(1+q_0+q_1)}\exp{\left(-q_1\frac{z}{1+z}\right)},
\label{hub1}
\ee

\be
Model ~ II.~~~~~H(z)=H_0(1+z)^{(1+q_0)}.(1+z^2)^{q_1/2},
\label{hub2}
\ee

\be
Model ~ III.~~~~~H(z)=H_0(1+z)^{(1+q_0+q_1)}\exp{\left[\frac{q_1}{2}\left(\frac{1}{(1+z)^2}-1\right)\right]}.
\label{hub3}
\ee

It is interesting to note that the parameter $q_0$ represents the present value of deceleration parameter in these parameterizations of $q(z)$. But the parameter $q_1$ in the expressions of $q(z)$ are not equivalent. For a better understanding, $q_1$ can be replaced by expressing in terms of $q_0$ and the redshift of transition from decelerated to accelerated phase of expansion ($z_t$). The parameter $q_1$ is related to $q_0$ and $z_t$ for these models as, Model I. $q_1=-q_0\frac{(1+z_t)}{z_t}$, Model II. $q_1=-q_0\frac{1+z_t^2}{z_t(1+z_t)}$, and Model III. $q_1=-q_0\frac{(1+z_t)^2}{(1+z_t)^2-1}$. Thus the transition redshift $z_t$ has been introduced as a model parameter in the present context. Statistical analysis of the models are discussed in the following section. The parameters which are constrained in the present context, are $h_0=H_0/100$ km s$^{-1}$Mpc$^{-1}$, the present value of deceleration parameter $q_0$ and the transition redshift $z_t$. The evolution of different cosmological quantities are also studied for the present kinematical models.

\section{Statistical analysis and observational constraints on the models}
\label{Stat}

An indispensable part of a reconstruction is the statistical analysis of the model based on observational data set. In the present context, statistical analysis has been carried out using different observational data sets, namely the supernova distance modulus data, observational measurements of Hubble parameter and local measurement of Hubble constant. The supernovae distance modulus measurements data from the Joint Light-curve Analysis (JLA) \cite{jla} has been utilized in the present context. The observational measurements of Hubble parameter (OHD) in the redshift range $0.07<z<2.36$  that include Cosmic Chronemeter measurements \cite{ohdcc}, measurement of Hubble parameter from baryon acoustic oscillation galaxy distributions \cite{ohdbao}, measurement of Hubble parameter from Lyman-$\alpha$ forest \cite{ohdLya} are also included in the analysis. We have also incorporated the SH0ES measurement of presentday Hubble expansion  rate $H_0=73.52\pm 1.62$ km s$^{-1}$Mpc$^{-1}$ (R18) \cite{Riess:2018byc}.

\begin{table}[htb]
\caption{{\small Parameter values, obtained in the statistical analysis with different combinations of the data sets. The mean values of the parameters along with the associated 1$\sigma$ uncertainties are given. The corresponding $\chi^2_{min}$ values are also shown.}}
\begin{center}
\resizebox{0.82\textwidth}{!}{  
\begin{tabular}{c |c |c c c c |c   } 
  \hline
 \hline
Model & Data Sets & $h_0$ &  $q_{0}$  &  $z_{t}$ & $q_1$ & $\chi^2_{min}$\\ 
 \hline
&  OHD+JLA & $0.700\pm 0.008$ & $-0.597\pm 0.064$ &   $0.809^{+0.058}_{-0.050}$ & $1.335_{-0.187}^{+0.197}$ & 49.87\\ 
I & OHD+JLA+R18 & $0.707\pm 0.007$ & $-0.640\pm 0.060$ &   $0.818^{+0.056}_{-0.047}$ & $1.422_{-0.178}^{+0.186}$ & 53.62\\

 \hline
& OHD+JLA & $0.699\pm 0.008$ & $-0.538\pm 0.058$ &  $0.740^{+0.046}_{-0.037}$ & $0.647_{-0.096}^{+0.097}$ & 49.61\\ 
II & OHD+JLA+R18 & $0.705\pm 0.007$ & $-0.576\pm 0.054$ &   $0.747^{+0.044}_{-0.037}$ & $0.688_{-0.089}^{+0.105}$ & 53.61 \\

 \hline  
& OHD+JLA & $0.702\pm 0.008$ & $-0.690\pm 0.075$ &  $0.726^{+0.061}_{-0.049}$ & $1.039_{-0.144}^{+0.148}$ & 50.08\\
III & OHD+JLA+R18 & $0.709\pm 0.007$ & $-0.736\pm 0.069$ &  $0.735^{+0.058}_{-0.047}$ & $1.102_{-0.134}^{+0.138}$ & 53.32\\ 

 \hline
  \hline
\end{tabular}
}
\end{center}
\label{table_results}
\end{table}

Bayesian statistical inference is adopted here to estimate the posterior probability distribution of the parameters. Bayesian statistical inference suggests that the posterior probability distribution of a parameter is proportional to the distribution of likelihood and the prior information. Uniform prior distributions are adopted in the present analysis. The likelihood incorporates the observation data in the analysis.  The parameter values are estimated by Markov chain Monte Carlo (MCMC) analysis using the {\footnotesize PYTHON} implication of MCMC sampler {\footnotesize EMCEE}, introduced by Goodman and Wear \cite{emcee1} and by Foreman-Mackey et al. \cite{emcee2}. The parameter space consist of the present Hubble parameter, scaled by $100$km s$^{-1}$Mpc$^{-1}$ ($h_0$), the present value of the deceleration parameter ($q_0$) and the transition redshift $z_t$. The parameters values, obtained in the present analysis are given in table \ref{table_results}. The statistical analysis has been carried out for two combinations of  the data sets, OHD+JLA and OHD+JLA+R18. It is found that the addition of the local measurement of Hubble constant (R18) increases the value of present Hubble parameter $h_0$. The other two parameters are get slightly changed with the addition of R18 measurement. The present deceleration parameter $q_0$ value decreases  and the transition redshift  $z_t$ slightly increases with the addition of R18 data. The negative value of the parameter $q_0$ ensures the present accelerated expansion. The transition redshift $z_t$ is found to be $z_t<1$ for all three kinematic models. So the present kinematical models indicate that the transition from decelerated to accelerated phase occurred in the recent past. The transition redshift predicted in Model I is slightly higher that value predicted in other two cases, but all the values are consistent to each other within 1$\sigma$ uncertainty.  The two dimensional (2D) confidence contours and the posterior probability distribution of the parameters for these models are shown in figure \ref{cont_M1}, \ref{cont_M2} and \ref{cont_M3}. The parameter $h_0$ and $q_0$ has a sharp negative correlation. On the other hand, $z_t$ is found to be very weekly correlated with $h_0$ and $q_0$. The profile of correlations amongs the parameters are very similar in all these three kinematic model.

The minimum values of $\chi^2$ ($\chi^2_{min}$) for present analysis are also shown in table \ref{table_results}. The $\chi^2_{min}$ values are very close for all three models. On the other hand, same sets of data and same number of parameters are involved in the statistical analysis of the models. The $\chi^2_{min}$ values reveal that present kinematic models have almost equal statistical preferences based on the MCMC analysis for parameter fitting.

\begin{figure}[tb]
\begin{center}
\includegraphics[angle=0, width=0.32\textwidth]{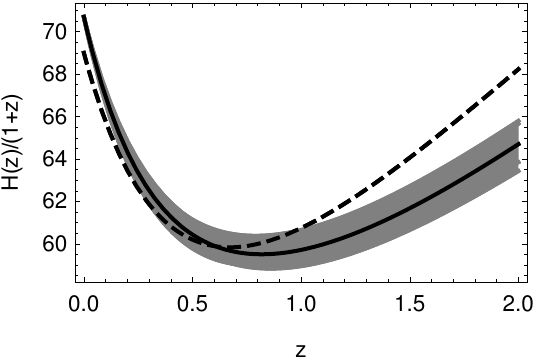}
\includegraphics[angle=0, width=0.32\textwidth]{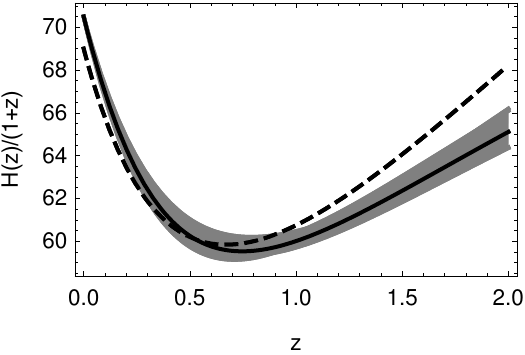}
\includegraphics[angle=0, width=0.32\textwidth]{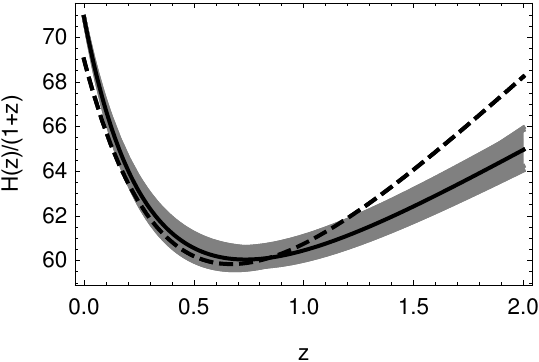}\\
\end{center}
\caption{{\small Plots of $H(z)/(1+z)$ for the reconstructed models (black solid curves) and the $\Lambda$CDM model (dashed curves). The left panel is for Model I, middle panel is for Model II and  right panel is for Model III. The values of the parameter are kept at the values obtained in the statistical analysis combining OHD+JAL+R18. The $\Lambda$CDM curve is obtained for the parameter values obtained in Planck2018 \cite{Aghanim:2018eyx}.}}
\label{Hub}
\end{figure}
\begin{figure}[tb]
\begin{center}
\includegraphics[angle=0, width=0.32\textwidth]{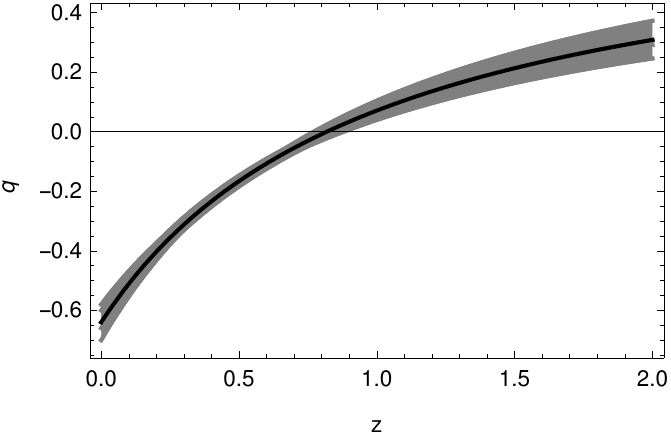}
\includegraphics[angle=0, width=0.32\textwidth]{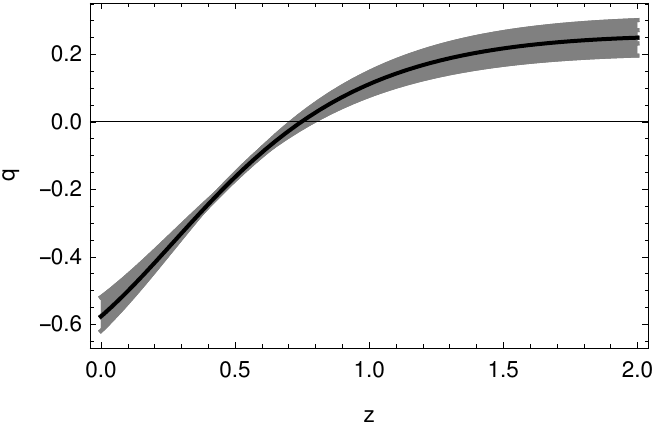}
\includegraphics[angle=0, width=0.32\textwidth]{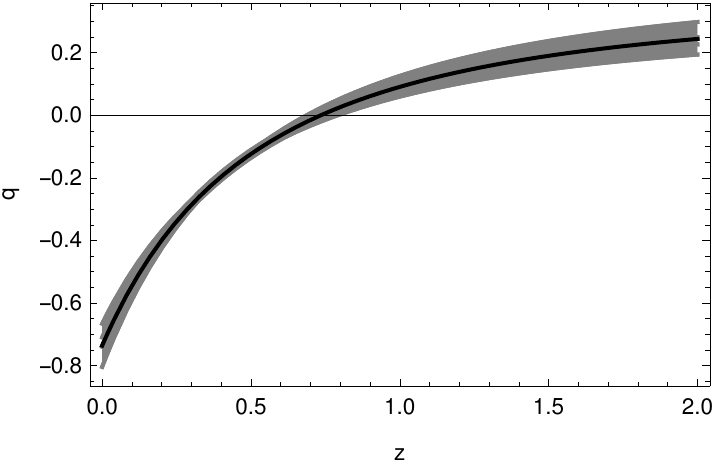}\\
\includegraphics[angle=0, width=0.32\textwidth]{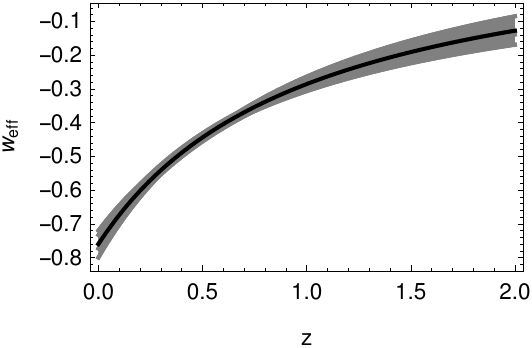}
\includegraphics[angle=0, width=0.32\textwidth]{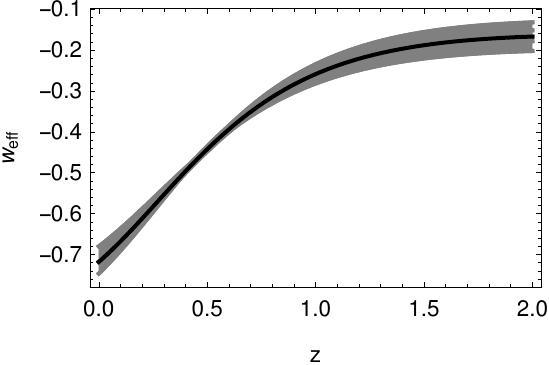}
\includegraphics[angle=0, width=0.32\textwidth]{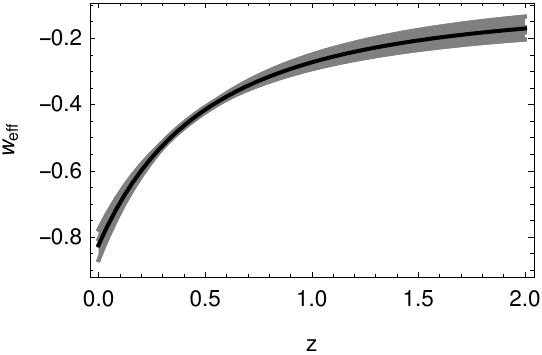}
\end{center}
\caption{{\small Plots of deceleration parameter (upper panels) and effective equation of state parameter (lower panels) for the reconstructed models. The left panels are for Model I, middle panels are for Model II and  right panel are for Model III. The best fit curve and the associated 1$\sigma$ confidence regions are shown.}}
\label{q_weff}
\end{figure}

\begin{figure}[h]
\begin{center}
\includegraphics[angle=0, width=0.32\textwidth]{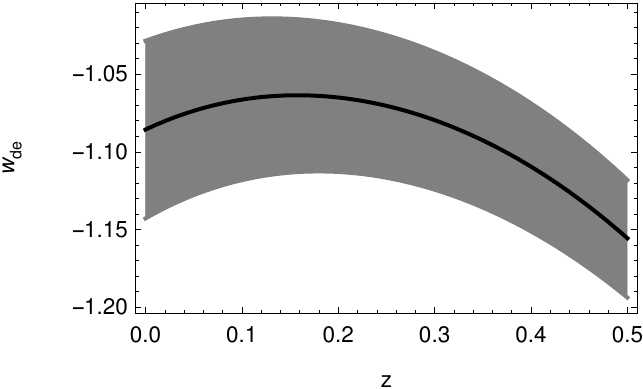}
\includegraphics[angle=0, width=0.32\textwidth]{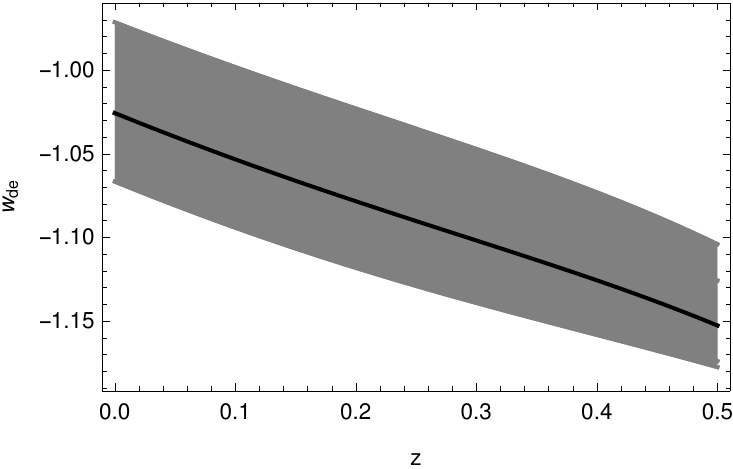}
\includegraphics[angle=0, width=0.32\textwidth]{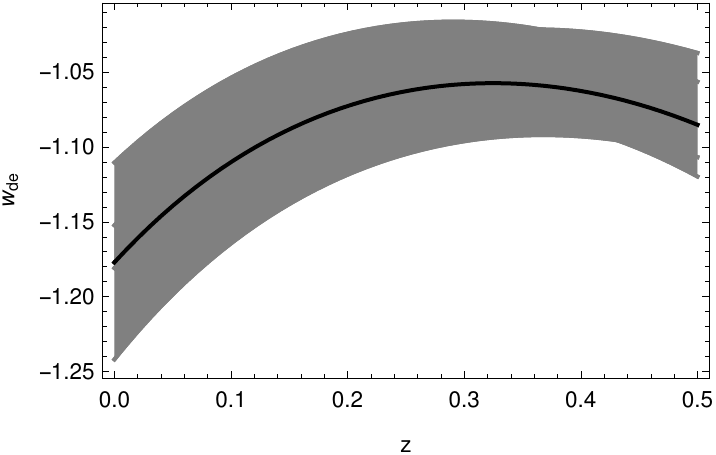}
\end{center}
\caption{{\small Plots show the nature of dark energy equation of state parameter ($w_{de}$) for the reconstructed kinematic models. The best fit curve and the 1$\sigma$ confidence regions are  shown. The left panel is for Model I, middle panel is for Model II and  right panel is for Model III.}}
\label{wde}
\end{figure}

Evolution of $H(z)/(1+z)$ are shown in figure \ref{Hub}. The curves obtained for the present kinematical models and the $\Lambda$CDM cosmology are shown. There is a deviation from the $\Lambda$CDM curve at higher redshift. The deceleration parameter at the best fit and 1$\sigma$ confidence region are shown in the upper panels of figure \ref{q_weff}. As already mentioned, the reconstructed kinematical quantities are independent of any assumption of the dynamical nature of the components in the energy budget of the universe and also do not depend on the undertaken gravity theory. But in the kinematic approach, we can not constrain the nature and evolution of dark energy directly. The dark energy properties can only be studied if we adopt the assumptions GR in the context of kinematical reconstruction.  In the present context, the dynamical nature of different components in the energy budget of the universe are studied under the regime of GR. The effective equation of state parameter of the total fluid content of the universe is defined as,
\be
w_{eff}=\frac{p_{tot}}{\rho_{tot}}.
\ee
The total energy density ($\rho_{tot}$) and the total effective pressure ($p_{tot}$) are connected to the expansion rate by the following relations,
\be
\frac{\rho_{tot}}{\rho_{c0}}=\frac{H^2(z)}{H^2_0},
\ee 
\be
\frac{p_{tot}}{\rho_{c0}}=-\frac{H^2(z)}{H^2_0}+\frac{2}{3}\frac{(1+z)H(z)H'(z)}{H^2_0}.
\ee
The $\rho_{c0}$ is the present critical density, defined as $\rho_{c0}=3H^2_0/8\pi G$. Thus the effective equation of state can be studied for the present kinemattical models under the regime of GR. In the lower panels of figure \ref{q_weff}, the evolution of $w_{eff}(z)$ are shown. It indicates toward an effective negative pressurelike contribution of the fluid content of the universe at the low redshift regime. At high redshift, the value of $w_{eff}$ rolls towards zero, indicating a dust matter dominated dynamics. Similarly the nature of dark energy equation of state can be studied. The additional assumption required for that is regarding the conservation of different components in the energy budget. In the present context, we assume that the dark energy and dark matter components are independently conserved. The contribution of radiation energy density can be neglected at low redshift regime. Thus in a spatially flat universe, the dark energy density can be expressed as, $\rho_{de}(z)/\rho_{c0}=(H(z)/H_0)^2-\Omega_{m0}(1+z)^3$. The present matter density parameter $\Omega_{m0}$ is defined as, $\rho_{m0}/\rho_{c0}$. Only the dark energy contributes to the total fluid pressure, thus $p_{de}=p_{tot}$. The dark energy equation of state parameter ($w_{de}=p_{de}/\rho_{de}$) for the reconstructed kinematical models are shown in figure  \ref{wde}. It reveals a phantom nature of dark energy in the present epoch.

\section{Dark energy diagnostics and statefinder hierarchy}
\label{statefinder}

\begin{figure}[tb]
\begin{center}
\includegraphics[angle=0, width=0.3\textwidth]{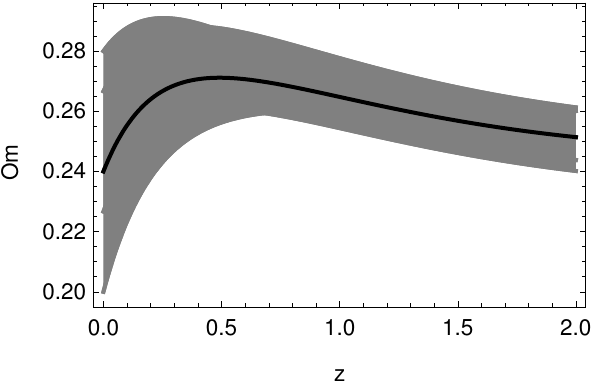}
\includegraphics[angle=0, width=0.3\textwidth]{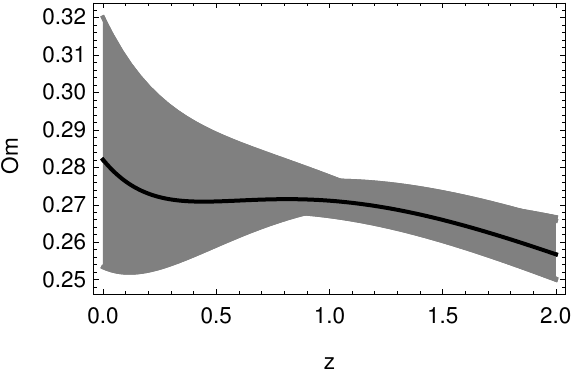}
\includegraphics[angle=0, width=0.3\textwidth]{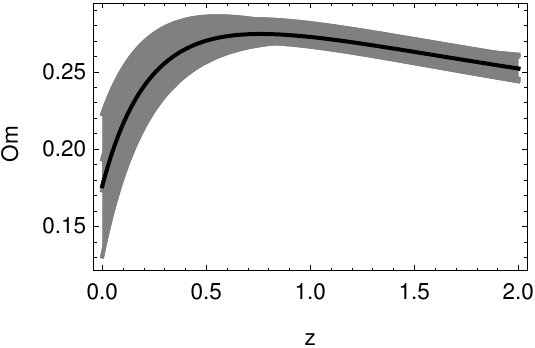}
\includegraphics[angle=0, width=0.29\textwidth]{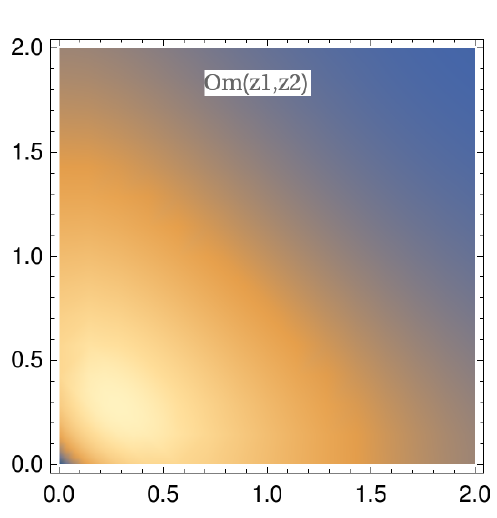}
\includegraphics[angle=0, width=0.29\textwidth]{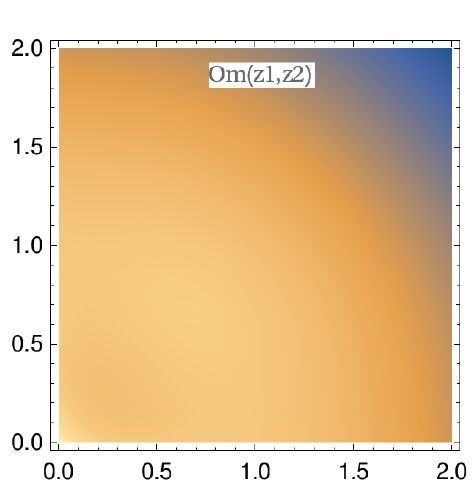}
\includegraphics[angle=0, width=0.29\textwidth]{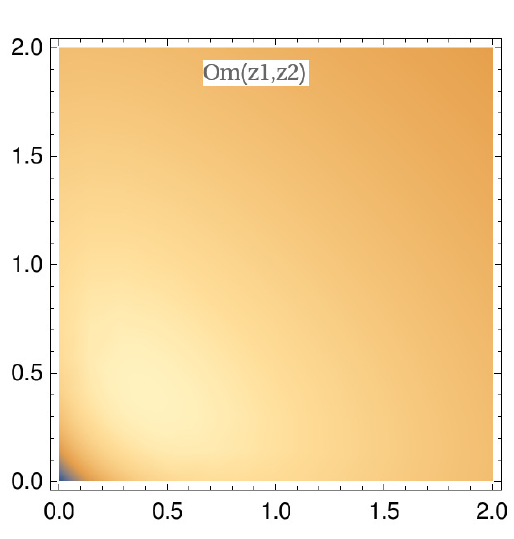}
\end{center}
\caption{{\small Plots of Om(z) (upper panels) and the colour coded presentation of two point functions $Om(z_i,z_j)$ (lower panels). The bestfit curve and the associated 1$\sigma$ confidence regions are shown. The left panels are for Model I, middle panels are for Models II and right panels are for Model III.}}
\label{Om_plots}
\end{figure}

\begin{figure}[tb]
\begin{center}
\includegraphics[angle=0, width=0.3\textwidth]{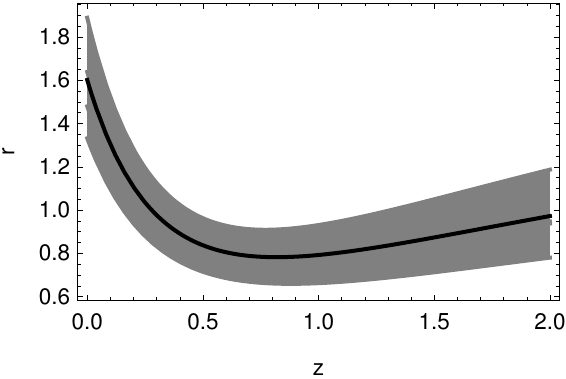}
\includegraphics[angle=0, width=0.3\textwidth]{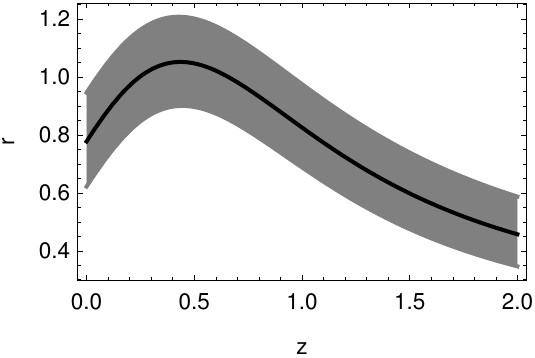}
\includegraphics[angle=0, width=0.3\textwidth]{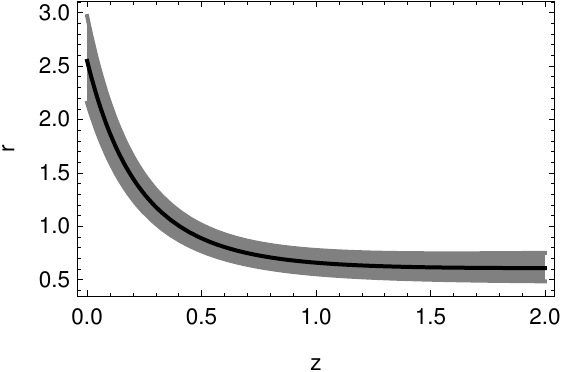}\\
\includegraphics[angle=0, width=0.3\textwidth]{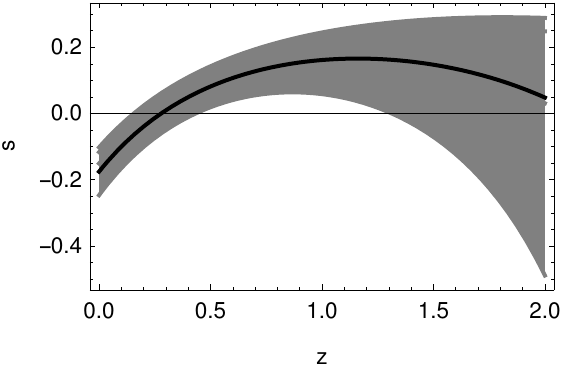}
\includegraphics[angle=0, width=0.3\textwidth]{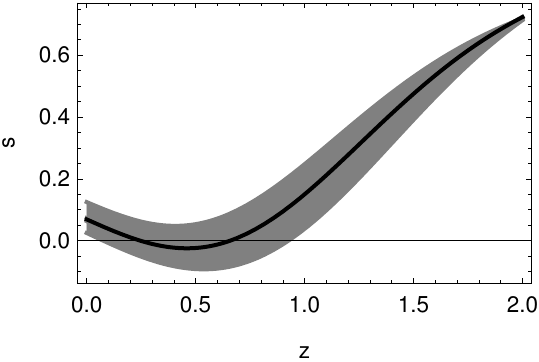}
\includegraphics[angle=0, width=0.3\textwidth]{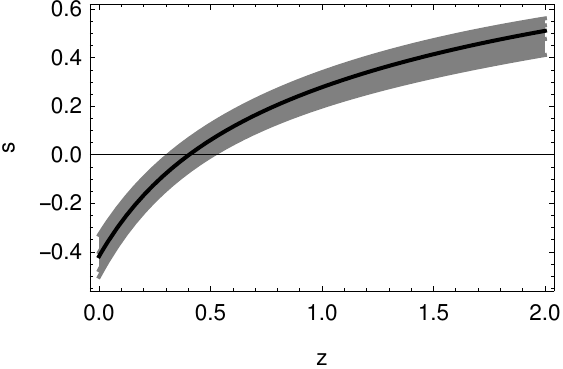}
\end{center}
\caption{{\small Evolution of state finder parameters $r$ (upper panels) and $s$ (lower panels) are shown. The bestfit curves and the associated 1$\sigma$ confidence regions are presented. The left panels are for Model I, middle panels are for Models II and right panels are for Model III.}}
\label{S_plots}
\end{figure}
\begin{figure}[t]
\begin{center}
\includegraphics[angle=0, width=0.45\textwidth]{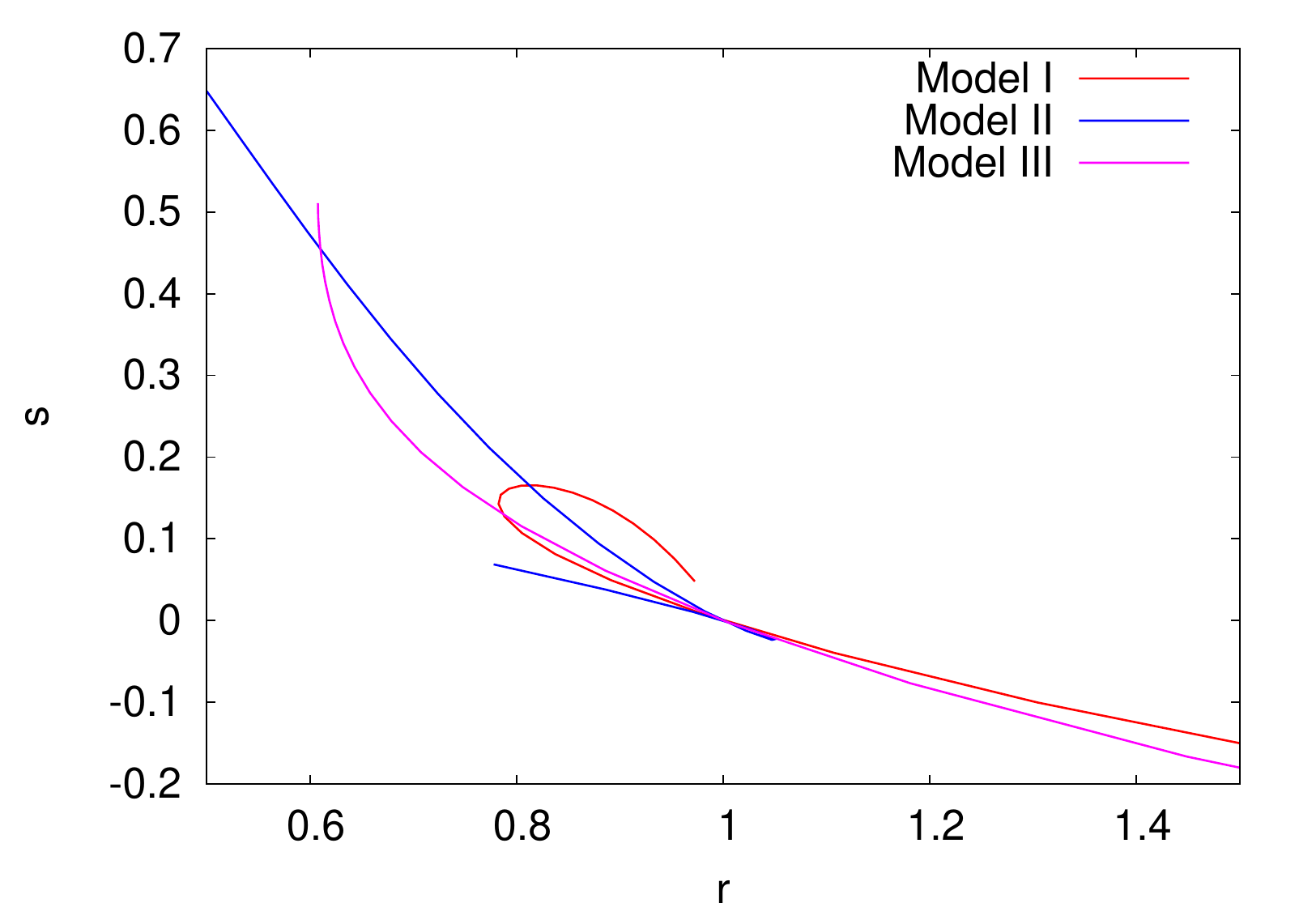}
\end{center}
\caption{{\small Statefinder phasespace diagram for the reconstructed models. Curves are obtained for the bestfit values of the parameters. The statefinder phase space is constructed with the parameters $r$ and $s$. The $\Lambda$CDM model is represented by the (1,0) point on the phase space.}}
\label{StatePhasespace_plots}
\end{figure}

In this section, the two different diagnostics of dark energy, namely the {\it Om}-diagnostic and the {\it statefinder} diagnostics are discussed. 
The $Om(z)$, introduced by Sahni et al. \cite{Sahni:2008xx} and by Zunckel and Clarkson \cite{Zunckel:2008ti}, is defined as,
\be
Om(z)=\frac{E^2(z)-1}{(1+z)^3-1},
\label{Om}
\ee
where $E(z)=H(z)/H_0$. For $\Lambda$CDM cosmology, the value of $Om(z)$ remains constant which is equal to $\Omega_{m0}$, the present value of the matter density scaled by the present critical density. Evolution in the value of $Om(z)$ diagnostic carries the signature of deviation from $\Lambda$CDM. As the $Om$ only depends upon the expansion rate, it is easier to be determined from the present observations. The $Om$-diagnostics can also be defined as a two-point function \cite{Shafieloo2012},
\be
Om(z_i;z_j)=\frac{E^2(z_i)-E^2(z_j)}{(1+z_i)^3-(1+z_j)^3}.
\label{Om2}
\ee
In figure \ref{Om_plots}, the $Om(z)$ and the two point function $Om(z_i,z_j)$ are shown. The reconstructed Model I allows a constant value of $Om$ with the 1$\sigma$ confidence region indicating a consistency with $\Lambda$CDM. The two point function $Om(z_i,z_j)$, presented in colour coding is shown in the lower panel of figure \ref{Om_plots}. The two point function depicts a slight variation in its value for the present models.  

Another diagnostic of dark energy is the {\it statefinder}. As the statefinder parameters contain higher order derivatives of the expansion factor, it is more sensitive and efficient to break the degeneracy in the dark energy models. The statefinders are defined in terms of kinematical parameters of different orders. As already mentioned that the deceleration parameter is the second order kinematical parameter. Similarly the 3rd order dimensionless kinematical parameter, the jerk ($j$) is defined as,
\be
j=\frac{1}{aH^3}\frac{d^3a}{dt^3}.
\ee       
In $\Lambda$CDM cosmology, the kinematical parameters $q$ and $j$ are expressed as \cite{Arabsalmani:2011fz},
\be
-q=1-\frac{3}{2}\tilde{\Omega}_m;~~~~~~j=1.
\label{a2}
\ee
where $\tilde{\Omega}_m=\Omega_{m0}(1+z)^3/E^2(z)$. The statefinder parameters of 3rd order ($r,s$) are defined as,
\be
r=j; ~~~~~ s=\frac{r-1}{3(q-\frac{1}{2})}.
\ee
Statefinder parameters are defined such a way that the $\Lambda$CDM cosmology is represented by a single point on the statefinder parameter space. The higher order statefinders can also be constructed in the similar way. Thus it forms a hierarchy of statefinder parameters, discussed in \cite{Arabsalmani:2011fz}. In the present analysis, we have studied the 3rd order statefinders $r$ and $s$.  
The two kinematical parameter together $\{r,s\}$ form a null diagnostic of $\Lambda$CDM cosmology  as their values are $\{1,0\}$ in $\Lambda$CDM. The evolution of the statefinder parameters are studied for the present kinematical models (figure \ref{S_plots}). The models are found to be distinguishable form  the nature of the statefinders. Only the Model I is found to be consistent with $\Lambda$CDM at higer redshift though the present value is in 1$\sigma$ tension with that of $\Lambda$CDM. Other two models show significant deviation form $\Lambda$CDM in the nature of the statfinder parameters. Nature of the hierarchy of statefinder parameters in the context of dynamical and kinematical models of latetime acceleration has earlier been discussed in \cite{Mukherjee:2018oll,Arabsalmani:2011fz}.

Further the ($r,s$) statefinder phase space behaviour for these models are studied. The phasespace behaviour of the present kinematical models are depicted in figure \ref{StatePhasespace_plots}. It revels that the ($r,s$) phase space evolution breaks the degeneracy in the models. The $\Lambda$CDM cosmology is represented by a the (1,0) point on the phasespace. It is interesting to note form the phasespace diagram that the present models attain the corresponding $\Lambda$CDM point once in course of evolution. The present analysis clearly reveals that the analysis of statefinder parameters successfully break the degeneracy of the present kinematical models. The statefinder diagnostics and the hierarchy of statefinder parameters are important tools to break the degeneracy of different kinematical and dynamical models of late-time cosmology.

\section{Evolution of matter density contrast at linear level}
\label{deltam}

\begin{figure}[tb]
\begin{center}
\includegraphics[angle=0, width=0.3\textwidth]{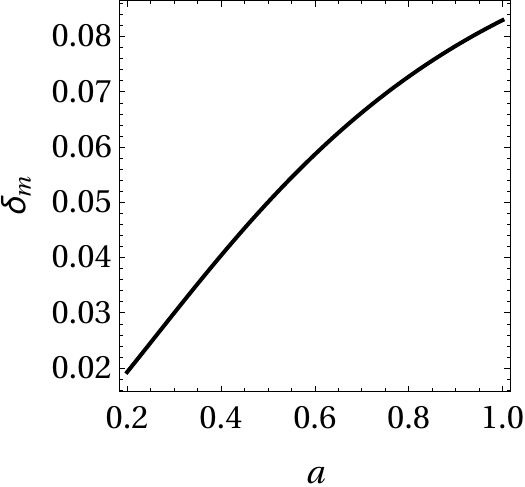}
\includegraphics[angle=0, width=0.34\textwidth]{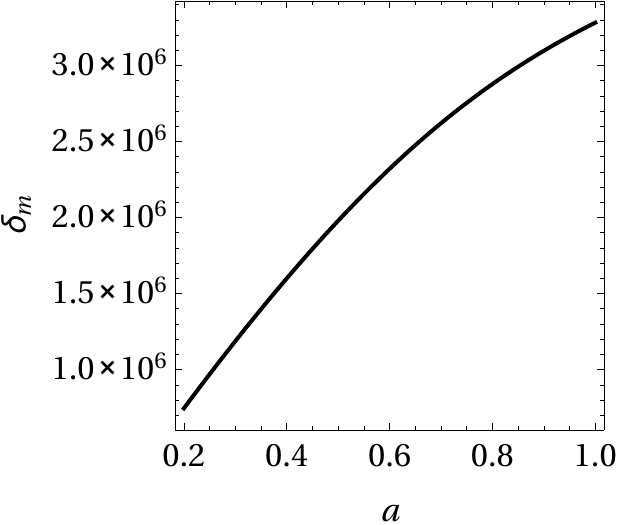}
\includegraphics[angle=0, width=0.29\textwidth]{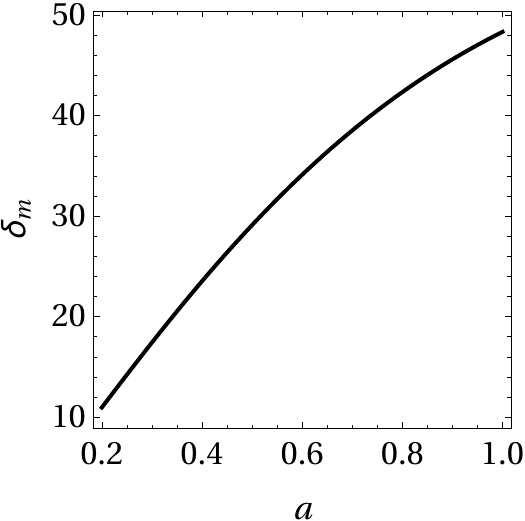}
\end{center}
\caption{{\small Evolution of the matter density contrast $\delta_m$ for the reconstructed kinematical models with the best fit values of the parameters. The left panel is for Model I, middle panel is for Model II and  right panel is for Model III. }}
\label{delm}
\end{figure}

\begin{figure}[h]
\begin{center}
\includegraphics[angle=0, width=0.3\textwidth]{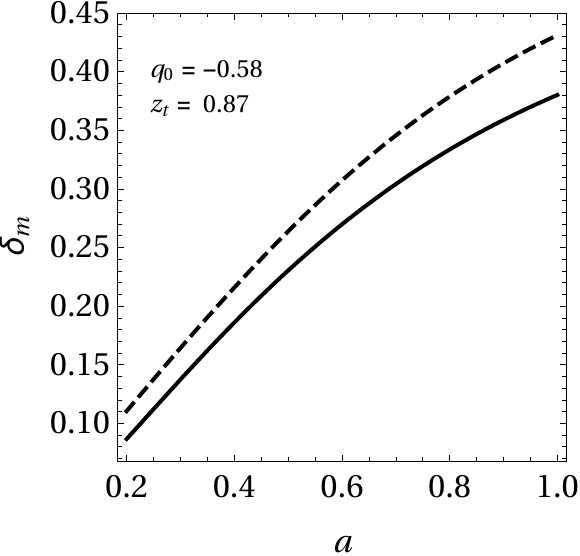}
\includegraphics[angle=0, width=0.3\textwidth]{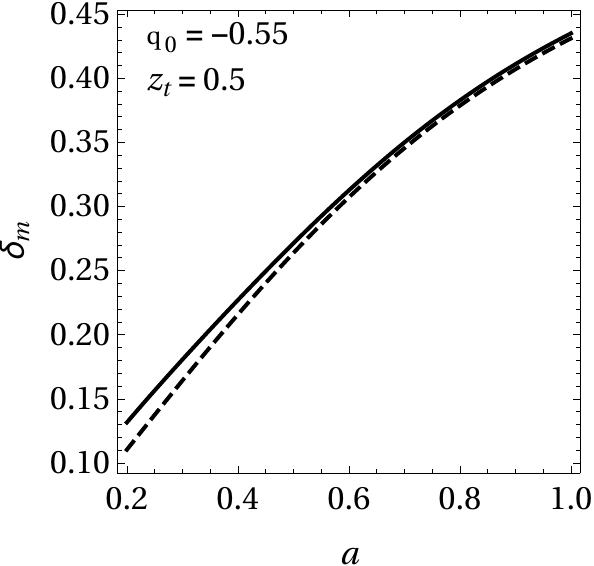}
\includegraphics[angle=0, width=0.3\textwidth]{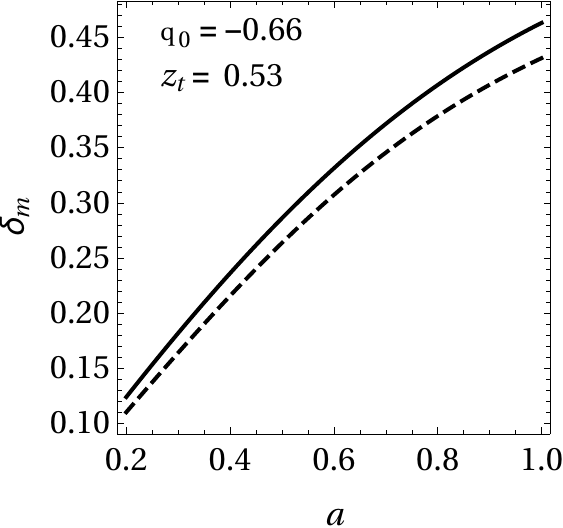}
\end{center}
\caption{{\small The evolution of matter density contrast $\delta_m$ (solid curves) for the reconstructed kinemtical models with adjusted values of the parameters $(q_0,z_t)$. The $\delta_m$ for $\Lambda$CDM cosmology (dashed curves) are also shown. The left panel is for Model I, middle panel is for Model II and  right panel is for Model III. The values of the parameters ($q_0,z_t$) for the $\delta_m$ (solid curves) are mentioned.}}
\label{delm_LCDM}
\end{figure}

The matter density contrast is defined as, $\delta_m=\delta\rho_m/\rho_m$, where $\rho_m$ is the homogeneous matter density at the background and $\delta\rho_m$  is the deviation from the homogeneous matter density. Due to the gravitational attraction, the matter over-density grows by accumulating mass from the surrounding. The evolution of $\delta_m$ at linear level is governed by the following equation,
\be
\ddot{\delta}_m+2H\dot{\delta}_m=4\pi G\rho_m\delta_m.
\label{delm_eq}
\ee
where the overhead dots denote the differentiations with respect to cosmic time. The evolution of $\delta_m$ becomes nonliner near the gravitational collapse of the overdense region. Both the linear and non-linear evolution is deeply effected by the background expansion rate. In the present analysis, the linear evolution of $\delta_m$ is studied for the reconstructed kinematical models. Equation (\ref{delm_eq}) can be safely assumed as the observables in the present analysis are at the scale of acoustic sound horizon. Inhomogeneity in dark energy distribution contributes at very large scale only and hence can be neglected at the present analysis.
Taking the scale factor `a' as the argument of differentiation, equation (\ref{delm_eq}) can be written as.

\be
\frac{d^2\delta_m}{da^2}+\left(\frac{3}{a} +\frac{1}{E}\frac{dE}{da}\right)\frac{d\delta_m}{da}=\frac{3}{2}\frac{\Omega_{m0}}{a^5E^2}.
\label{delm_eqa}
\ee 

where $E(z)=H(z)/H_0$. The Hubble expansion (equation (\ref{hub1}) to (\ref{hub3})) rate obtained for the present kinematical models are utilized.  Equation (\ref{delm_eqa}) is numerically studied for the present reconstructed models taking the scale factor as the argument of differentiations instead of time. The initial conditions are fixed at the scale factor $a=0.001$, which is close to the era of cosmic microwave background. The initial values are fixed as, $\delta_m(a_i)=0.001$ and $\dot{\delta}_m(a_i)=0$. The evolution of $\delta_m$ for the best fit values of the parameter $q_0$ and $z_t$ for the present kinematical models are shown in figure \ref{delm}.

As the present models are obtained from the kinematical parameter, it is important for these models to produce the evolution of cosmological perturbations consistent to that of cosmological standard model. The linear evolution of matter overdensity is emphasized in this context. The values of the kinematical parameters ($q_0,z_t$) are adjusted by hand to have the $\delta_m(a)$ which are close to the corresponding $\Lambda$CDM curve of $\delta_m$. In figure \ref{delm_LCDM}, the $\delta_m(a)$ curves for the adjusted values of ($q_0,z_t$) are shown along with the corresponding $\Lambda$CDm curve. The values, fixed for $q_0$ and $z_t$ are also mentioned there in th plots. The value of present matter density parameter $\Omega_{m0}$ is fixed as $\Omega_{m0}=0.3$. It is found that the values of $q_0$ and $z_t$, required to produce $\Lambda$CDM like $\delta_m$ curve, remains with the 2$\sigma$ confidence region only for the Model I.

\section{Conclusion}
\label{conclu}

The present works deals with the kinematic models of late-time cosmology. The idea is to start with some phenomenological parameterizations  of any kinematic quantity. We have utilized the parameterizations of the deceleration parameter. The kinematical parameters $h_0$, $q_0$, $z_t$ are constrained for the reconstructed models. These parameter values, obtained in the statistical analysis of the models using cosmological data sets, are found to be consistent for different models. It worth mention at this point that the transition redshift $z_t$ is dependent on the expression of the deceleration parameter, thus $z_t$ can have highly different values for different assumptions about the deceleration parameter. In the present models, the values of $z_t$ are found to be consistent at 1$\sigma$ level.

The dynamics of the universe has been investigated for the reconstructed kinematical models under the regime of GR. The evolution of the effective equation of state of the total fluid content of the universe has been studied. It is apparently clear from the nature of the effective equation of state that the present universe is dominated by the component which has a negative pressuelike contribution and at the high redshift, the universe was dominated by pressureless component. With the assumption of independent conservation of dark energy and dark matter components in a spatially flat universe, the present nature of dark energy equation of state is reconstructed through the kinematical models. The present nature of dark energy equation of state is found to be in phantom regime. 

The statedinder diagnostic has been emphasized in the present context. It is found that the statefinders successfully break the degeneracy of the models that prevail in the nature of Hubble and deceleration parameter and also in the nature of dark energy equation of state. The statefinders also indicate the deviation of the model from $\Lambda$CDM. Only the reconstructed model I allows the corresponding $\Lambda$CDM value of the state finder within the 1$\sigma$ confidence region at high redshift. The statefinder phase space diagrams are also distinct for the models. Interestingly, the phase space curves of the models pass through the corresponding $\Lambda$CDM point on the phase space in course of evolution.

Study of cosmological evolution using kinematic approach are already there in literature. In the present context, some of  the simple expressions of deceleration parameter are adopted for the study. In the present analysis, some new aspects are emphasized for  kinematic approach of cosmological reconstruction. The redshift of transition is introduced as a free parameter in the analysis. The correlation of transition redshift with other parameters are investigated. It is found the correlation of transition redshift with other parameters is very less. The value of transition redshift is obtained to be $z_t<1$, indicating a recent transition from decelerated to accelerated phase of expansion. A kinematic analysis with supernovae data has been carried out by Riess {\it et al} \cite{Riess:2004nr} where the recent transition from decelerated to accelerated phase of expansion was first revealed. In the present study, it is shown that the transition redshift does not varies significantly with variation of the cosmological model. Similar assumptions of deceleration parameter have been adopted in the study of cosmological dynamics by Gong and Wang \cite{Gong:2006tx,Gong:2006gs}, where they have used the earlier sample of type Ia supernovae distance modulus. With the increasing number of data points in the type Ia supernovae data compilation and the additions of the measurements of Hubble parameter at different redshifts put tighter constraints on the model parameters, namely the present deceleration parameter $q_0$ and the transition redshift $z_t$. Further in the present work, the nature of state finder parameters are also emphasized and thus the degeneracy in the models are broken. The evolution of liner matter perturbation is also studied in the present context. Mamon and Bamba \cite{Mamon:2018dxf} have studied the observational constraints on cosmological jerk parameter for different parametrizations of deceleration parameter, where different models are obtained from a general assumption about the deceleration parameter. Thus they obtained similar type of evolutions of jerk parameter for the models. But in the present study, the parametric forms of $q(z)$ are hardly connected through any general assumption of $q(z)$. It is found in the present analysis that though the evolution of $H(z)$ and $q(z)$ are degenerate, these models produce different behaviours of the jerk parameter and the statefinders. The study of dark energy properties for kinematical models is also a new aspect of the present work. In the present study, the nature of dark energy equation of state for the present kinematical model are emphasized. A GR assumption is required to investigate the dark energy propertied. Independent conservation of different components in the energy budget of the universe is assumed.

Besides the background, the evolution of matter perturbation at linear level has also been emphasized. Though it is totally based on GR assumptions,   it is important to study the viability of any model which is consistent at background level. It is another new aspect of the present study.  The evolution of matter density contrast for the present models are studied totally based on general relativistic equation (equation \ref{delm_eq}). The reconstructed Mode I is found to be consistent with the $\Lambda$CDM cosmology at linear level of matter perturbation within 1$\sigma$ confidence region. For other two models, the corresponding $\Lambda$CDM evolution of $\delta_m$ remain out of 2$\sigma$ confidence region. Thus it clearly makes reconstructed Model I preferable over Model II and Model III. It is important to note at these point that the evolution of matter density contrast is highly sensitive to the values of the kinematical parameters. But the present study of matter density perturbation for the present kinematical models hardly provide any information about the nature of dark energy.

The values of  Hubble constant, deceleration parameter and the transition redshift, obtained in the present analysis, are found to be totally consistent with the constraints on these parameters in different dark energy models. So, present kinematic approach indicates toward the dark energy model of cosmic acceleration. The energy budget of the present universe is dominated by dark energy and in the recent past it was dominated by dark matter.

\vskip 1.50 cm

\section*{Acknowledgment}
{\small The author would like to acknowledge the financial support from the Science and Engineering Research Board (SERB), Department of Science and Technology, Government of India through National Post-Doctoral Fellowship (NPDF, File no. PDF/2018/001859). The author would like to thank Prof. Anjan A. Sen for useful discussions.}

\vskip 0.50 cm


\end{document}